\documentclass[11pt,twoside]{article}
\usepackage{asp2010}

\resetcounters

\begin{document}

\title{HELGA: The Herschel Exploitation of the Local Galaxy Andromeda. Sub-mm morphology and dust properties.}
\author{Jacopo Fritz,$^1$ Matthew~W.~L. Smith,$^2$ Jason Kirk,$^{3,2}$\\ 
and the HELGA collaboration.
\affil{$^1$Sterrenkundig Observatorium, Universiteit Gent, Krijgslaan 281 S9, B-9000 Gent, Belgium; jacopo.fritz@UGent.be}
\affil{$^2$School of Physics \& Astronomy, Cardiff University, The Parade, Cardiff CF24 3AA, UK}
\affil{$^3$Jeremiah Horrocks Institute, University of Central Lancashire, Preston PR1 2HE, UK}}

\begin{abstract}
The results from a large-field Far-Infrared (FIR) and sub-millimeter (sub-mm) survey of our neighbor galaxy M31 are presented. We have obtained {\it Herschel} images of a $\sim5.5\times 2.5$ degree area centered on Andromeda. Using 21 cm atomic hydrogen maps, we are able to disentangle genuine emission from M31 from that for foreground Galactic cirrus, allowing us to recognize dusty structures out to $\sim 31$ kpc from the center. We first characterize the FIR and sub-mm morphology and then, by de-projecting {\it Herschel} maps and running an {\it ad--hoc} source extraction algorithm, we reconstruct the intrinsic morphology and the spatial distribution of the molecular complexes. Finally, we study the spatially resolved properties of the dust (temperature, emissivity, mass, etc.), by means of a pixel--by--pixel SED fitting approach. 
\end{abstract}

\section{Introduction}
We have observed the local galaxy Andromeda (M31), the nearest spiral galaxy in the Local Group, with {\it Herschel}. Simultaneous observations were conducted in five bands from 100 to 500 $\mu$m, of a $\sim5.5 \times 2.5$ degree area \citep{fritz12}. We study, for the first time at these wavebands, the characteristics of the extended dust emission, focusing on larger scales than ever done before for M31. Furthermore, thanks to the spatial resolution that the instruments onboard {\it Herschel} can reach, we are able to study the geometrical distribution and the properties of dust on a sub-kpc scale. 

Here, we summarize the results from a series of papers whose goal is the study of the morphology of the interstellar medium in Andromeda, with a particular focus on the physical properties of dust as a function of position and galactocentric distance.

\section{Dust morphology}
\begin{figure}
\begin{tabular}{ll}
\includegraphics[width=0.51\textwidth]{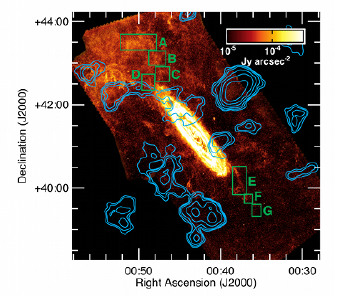} &
\includegraphics[width=0.45\textwidth]{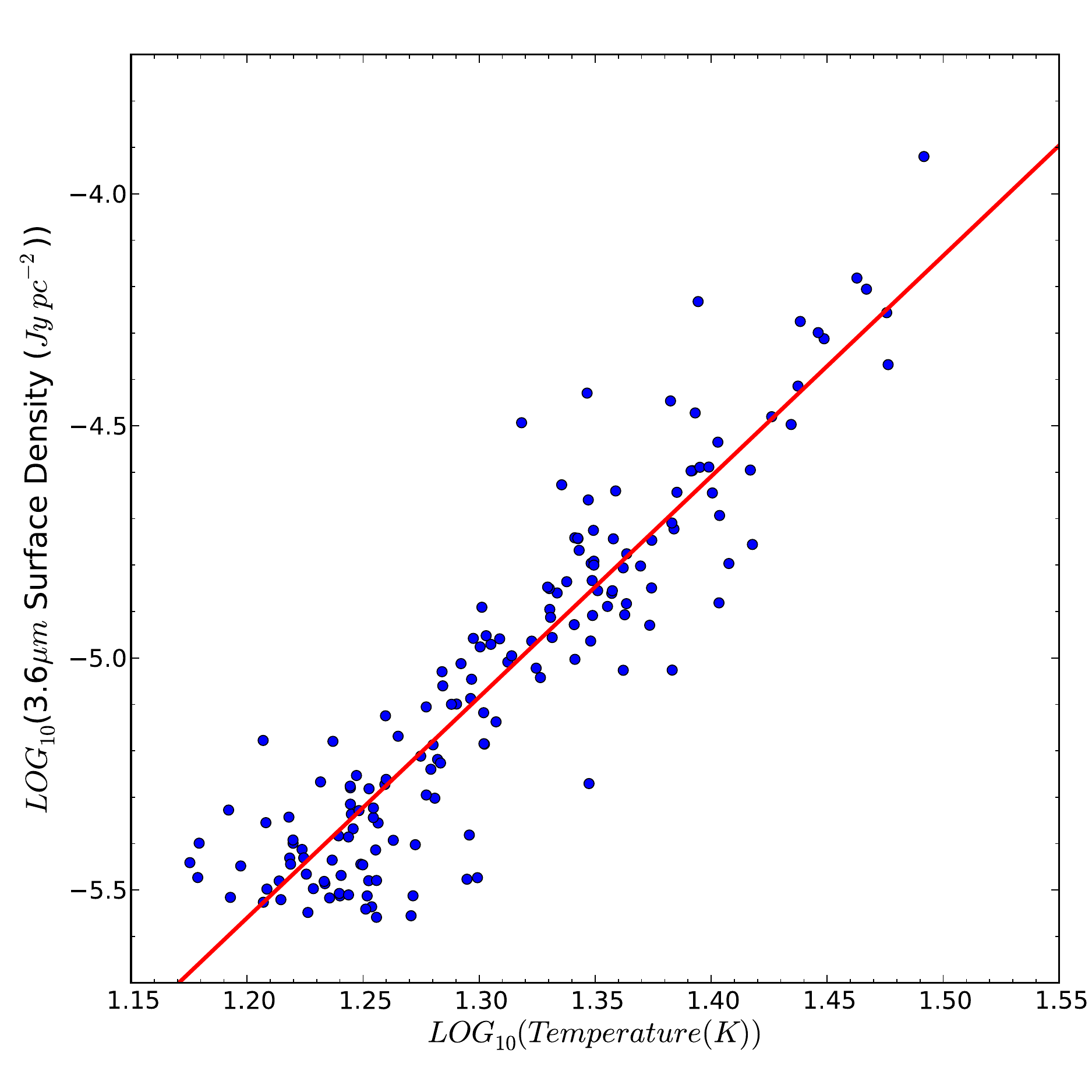}
\\
\end{tabular}
\caption{{\it Left-hand} panel: 250 $\mu$m map and high-velocity clouds (contours). Green squares highlight dusty structures possibly belonging to M31. {\it Right-hand} panel: correlation between the T$_D$ and 3.6 $\mu$m surface brightness for pixels within 3.1 kpc (see Sect.~4).}
\label{fig:fig1}
\end{figure}

In the left--hand panel of Fig.~\ref{fig:fig1}, we show a view of M31 at 250 $\mu$m. This wavelength gives the best compromise between sensitivity and spatial resolution. We have tried to identify structures belonging to M31 out to the largest distances, but this task turned out to be quite challenging due to the presence of a substantial amount of Galactic dust sitting on the line of sight between us and Andromeda. 

We have used the 21 cm hydrogen maps from \cite{thilker04} and \cite{braun09} to disentangle the dust emission coming from M31 from that of the Galactic cirrus. We exploit the physical link between gas and dust, and the fact that {\sc Hi} observations carry also the information on the gas velocity: this is used to tell which gas (and hence dust) component belongs to M31 and which one to the Milky Way. In this way, we manage to remove the Galactic cirrus emission from our data even though the north-east portion of our maps still remains quite confused, retaining some contamination. This only marginally affects the outskirts of M31, at {\it Herschel} wavebands.

We first looked for the presence of dust at large distances from M31 at the high-velocity clouds location \citep{thilker04}. A quick check of the most massive one, Davis' cloud, was enough to make us realize that our {\it Herschel} data are not sensitive enough for the detection of a possible dust component here.

Hence, we concentrated our search in the disk outskirts, specifically in the south--west region: the fact that there is no cirrus contamination makes the dust detection here quite clear, out to galactocentric  distances of $\sim 31$ kpc. Three ring-like structures are revealed, at distances of 21 (E), 26 (F) and 31 (G) kpc, respectively, from the nucleus. Even though quite faint, their detection is confirmed by the perfect coincidence with similar features observed in the 0-moment 21 cm map from \cite{braun09}. 

We have tried to check whether these features have a counterpart, at similar distances, on the opposite side of the galaxy. Due to the high degree of confusion of the north--east region, we are not able to tell whether this is the case or not, with the only exception of structure ``E'', the closest one, even though impossible to follow along all the disk of the galaxy. These newly discovered structures add to the already well--known morphological features of M31, such as the 10 kpc star forming ring and the 15 kpc ring \citep[already known for a while; see e.g.][]{haas98} and the more recently discovered 1.5 kpc, inner ring \citep{block06}.

Overall, we find evidence of a recurrent pattern in the morphological features of the dust in M31: rings, or more in general, arc-like structures are found at distances of 5 kpc each. If we also take into account a structure defined by \cite{baade63} as an arm-crossing region, located at $\sim 5$ kpc from the center, we note that  M31 has structures at radii of 5, 10, 15, 21, 26 and 31 kpc.  While the presence of these structures was never investigated before, the appearance of multiple rings in a galaxy was argued to be compatible with dissipative cloud collisions and weak bar perturbation, which would give rise to resonances, due to gravitational torques \citep{combes88,jungwiert96}.

\section{Molecular complexes}
We adopted a different, more systematic approach to the study of the ISM morphology in \cite{kirk13b}, where we de-projected our {\it Herschel} images and applied {\sc csar} \citep{kirk13a}, a source extraction algorithm suited for taking into account the hierarchical tree which builds up the arms and disks structure of the ISM in M31.

This yields a catalog of point sources which correspond to Giant Molecular Clouds (GMCs). Following a similar approach as in \cite{gordon06}, we fit 2 rings to the positions of the molecular clouds, ending up with a best fit radius of $10.52\pm 0.02$ and $15.50 \pm0.02$ kpc, respectively, reproducing the 10 and 15 kpc rings (see Sect.~2). We then fit 2 logarithmic spiral arms to the remaining points, letting the pitch angle as a free parameters. Such spiral arms are indeed suited to reproduce the GMCs dislocation, and they turn to have an angle of $8.9^\circ$.

Consistently with the findings by \cite{block06}, the best fit models for these two rings are displaced with respect to the optical centre of M31. It was argued that this displacement might be the result of a head--on collision with Andromeda's compact satellite M32, as dynamical simulations from the latter work tend indeed to confirm.

\section{Dust properties}
In \cite{smith12} we took advantage of {\it Herschel} high resolution data, and used PACS and SPIRE images to build the FIR spatially-resolved SED of Andromeda. We fit a modified black--body function to the SED of each of the pixels, limiting ourselves to those having a flux above $5\sigma$ in all bands. Despite this quite severe requirement, we are left with $\sim 4000$ pixels. 

Our fitting procedure considers 3 free parameters: the dust emissivity index $\beta$, the dust mass and its temperature (T$_D$). Like this, we assume that dust emits at one single temperature, while it is more likely that a range of temperatures, in the same position, would be a more realistic description. In spite of this simplification, we find a $\chi^2$ value less than 2.73 for more than 90\% of the pixels, showing that our model is an adequate description of the dust SED at a pixel scale.

The result of the fits are then be used to trace the properties of dust on a physical scale of about 140 pc, as a function of the location. Interestingly, we find the highest  values of T$_D$ ($\sim 30$ K) inside the bulge, which is dominated by old stars, while in the 10 kpc ring, where star formation is currently occurring, the temperature is constant, about 17 K only. Furthermore, if the radial profile is considered, we notice that the temperature monotonically decreases from the center out to $\sim3.1$ kpc, and then it assumes a slightly positive slope. This trend is mirrored by that of the emissivity index $\beta$ which increases out to the same distance, and then it decrease again towards the outskirts of the galaxy. As there is a well known degeneracy between dust temperature and the $\beta$ parameter, we used Montecarlo simulations to check whether these trends are spurious effect caused by this, and found that this is not the case.

The fact that the dust temperature is much higher in the bulge region, which is dominated by old stellar stars and where star formation activity is minimal if not absent at all, while it is significantly lower in the star--forming ring, rises a question about the dust heating sources and mechanism. It had already been argued that dust in some nearby galaxies and in M31 itself, might be significantly heated by a old stellar population. We investigated this, and we found a linear dependency of T$_D$ on the 3.6 $\mu$m emission in the inner 3.1 kpc region (Fig.\ref{fig:fig1}, right panel). As this band traces very well the presence of old stars, this correlation suggests that indeed the old bulge population is the one which is responsible for the dust heating in the central region.

We argue that different values of $\beta$ could maybe be due to differences in the composition of the dust, or substantial changes in the radiation field leading to changes in the grain size (from sputtering) or mantle loss.

\section{Conclusions}
We have used {\it Herschel} PACS and SPIRE observations to probe the dust properties and morphology in the nearby galaxy M31. We performed analysis both on a global and on a spatially resolved scale. We can summarize our results as follows: 
\begin{enumerate}
\setlength{\itemsep}{-2pt}
\item M31's IR morphology is dominated by 2 rings and 2 spiral arms;
\item there is a recurrent radial pattern in the presence of structures: arc-like features are detected at intervals of $\sim 5$ kpc, out to a galactocentric distance of $\sim 31$ kpc;
\item a variable emissivity index $\beta$ is required to fit the spatially resolved IR SED;
\item There is a transition region at a distance of $\sim 3.1$ kpc, marking different dust properties, or witnessing a change in the radiation field;
\item T$_D$ is the highest in the bulge (30 K), only reaching 17 K in the starforming ring;
\item the strong correlation between the 3.6 $\mu$m flux and T$_D$ in the inner regions strongly suggest that old stars in the bulge are the responsible for these ``high'' temperatures.
\end{enumerate}

\end{document}